\documentclass[aps,superscriptaddress,twocolumn,10pt]{revtex4-1}

\usepackage{epsfig}
\usepackage{amssymb}
\usepackage{amsmath}
\usepackage{amsfonts}
\usepackage{graphicx}
\usepackage{mathrsfs}
\usepackage[dvips]{color}
\usepackage{multirow}

\newcommand{\bsigma}{\boldsymbol{\sigma}}

\newcommand{\bcK}{\boldsymbol{\cK}}

\newcommand{\fn}{{\mathfrak{n}}}

\newcommand{\fz}{\mathfrak{z}}

\newcommand{\fK}{\mathfrak{K}}

\newcommand{\bb}{\mathbf{b}}

\newcommand{\bbe}{\mathbf{e}}

\newcommand{\bk}{\mathbf{k}}

\newcommand{\bt}{\mathbf{t}}

\newcommand{\bA}{\mathbf{A}}

\newcommand{\bH}{\mathbf{H}}
\newcommand{\bI}{\mathbf{I}}

\newcommand{\bM}{\mathbf{M}}

\newcommand{\bS}{\mathbf{S}}

\newcommand{\cH}{\mathcal{H}}

\newcommand{\cK}{\mathcal{K}}

\newcommand{\cN}{\mathcal{N}}

\newcommand{\cP}{\mathcal{P}}

\newcommand{\cT}{\mathcal{T}}

\newcommand{\be}{\begin{equation}}
\newcommand{\ee}{\end{equation}}
\newcommand{\bea}{\begin{eqnarray}}
\newcommand{\eea}{\end{eqnarray}}
\newcommand{\nn}{\nonumber}
\newcommand{\kt}{\rangle}
\newcommand{\br}{\langle}

\newcommand{\ed}{\end{document}}

\newcommand{\np}{\newpage}

\newcommand{\bi}{\begin{itemize}}
\newcommand{\ei}{\end{itemize}}

\newcommand{\bce}{\begin{center}}
\newcommand{\ece}{\end{center}}

\newcommand{\sF}{\mathscr{F}}

\newcommand{\sT}{\mathscr{T}}

\newcommand{\RE}{{\rm Re}}
\newcommand{\IM}{{\rm Im}}

\begin{document}


\title{A Class of Exactly Solvable Scattering Potentials in Two
Dimensions, Entangled State Pair Generation, and a Grazing Angle
Resonance Effect
}

\author{Farhang Loran}
\email[Email address: ]{loran@cc.iut.ac.ir}
\affiliation{Department of Physics, Isfahan University of Technology,
Isfahan 84156-83111, Iran}
\author{Ali~Mostafazadeh}
\email[Email address: ]{amostafazadeh@ku.edu.tr}
\affiliation{Departments of Mathematics and Physics, Ko\c{c} University,
34450 Sar{\i}yer, Istanbul, Turkey}

\begin{abstract}

We provide an exact solution of the scattering problem for the
potentials of the form $v(x,y)=\chi_a(x)[v_0(x)+ v_1(x)e^{i\alpha
y}]$, where
$\chi_a(x):=1$ for $x\in[0,a]$, $\chi_a(x):=0$ for $x\notin[0,a]$,
$v_j(x)$ are real or complex-valued functions, $\chi_a(x)v_0(x)$ is
an exactly solvable scattering potential in one dimension, and
$\alpha$ is a positive real parameter.
If $\alpha$ exceeds the wavenumber $k$ of the incident wave, the
scattered wave does not depend on the choice of $v_1(x)$. In
particular, $v(x,y)$ is invisible if $v_0(x)=0$ and $k<\alpha$. For
$k>\alpha$ and $v_1(x)\neq 0$, the scattered wave consists of a
finite number of coherent plane-wave pairs $\psi_n^\pm$ with
wavevector: $\bk_n=(\pm\sqrt{k^2-(n\alpha)^2},n\alpha)$, where
$n=0,1,2,\cdots<k/\alpha$. This generalizes to the scattering of
wavepackets and suggests means for generating quantum states with a
quantized component of momentum and pairs of states with an
entangled momentum. We examine a realization of these potentials in
terms of certain optical slabs. If $k=N\alpha$ for some positive
integer $N$, $\psi_N^\pm$ coalesce and their amplitude diverge. If
$k$ exceeds $N\alpha$ slightly, $\psi_N^\pm$ have a much larger
amplitude than $\psi_n^\pm$ with $n<N$. This marks a resonance
effect that arises for the scattered waves whose wavevector makes a
small angle with the faces of the slab.

\end{abstract}

\maketitle

Exactly solvable models have played a central role in our
understanding of the conceptual and practical aspects of quantum
mechanics  \cite{practial-QM}. They are, in particular,
indispensable for developing computational techniques that make the
study of realistic quantum systems possible. This is why a detailed
treatment of these models forms an integral part of teaching quantum
mechanics. In this respect, it is quite unfortunate that the
discovery of a new exactly solvable model is an extremely rare
event. The purpose of this article is to report such a discovery. In
particular, we introduce a class of complex potentials in two
dimensions whose scattering problem admits an exact and closed-form
solution in terms of the elementary functions.  We use this solution
to offer an exact description of the scattering of polarized
electromagnetic waves by certain optical slab systems. This reveals
a number of intriguing properties of these systems that make them of
immediate interest for engineering optical devices such as a source
for coherent plane-wave pairs, quantum states with entangled
momentum, states with a quantized momentum along one direction, and
directional multimode lasers.

An exclusive feature of complex potentials in one dimension is their
ability to display nonreciprocal reflection \cite{jpa-2014c}. An
extreme example is a unidirectionally invisible potential, which is
reflectionless from one direction and displays perfect transmission
from both directions
\cite{invisible1a,invisible1b,invisible1c,lin,invisible2,pra-2013a,horsley,longhi-epl}.
The principal example is \cite{lin,longhi-jpa-2011,uzdin,jones-2012}
    \be
    v(x)=\fz\,e^{2\pi i x/a}\chi_a(x)=\left\{
    \begin{array}{cc}
    \fz\,e^{2\pi i x/a} & {\rm for}~x \in[0,a],\\
    0 & {\rm otherwise}.\end{array}
    \right.,
    \label{1d-potential}
    \ee
where $\fz$ and $a$ are real parameters, $\chi_a(x):=1$ for
$x\in[0,a]$,  and $\chi_a(x):=0$ for $x\notin[0,a]$.
To a very good approximation, this potential is invisible from the
left for incident plane waves of wavenumber $k=\pi/a$ provided that
$|\fz|a^2\ll 1$, \cite{pra-2014a}.

Recently, we have developed a multidimensional transfer-matrix
formulation of scattering theory \cite{pra-2016} that allows for a
straightforward treatment of the delta-function potentials in two
and three-dimensions, and paves the way for a multidimensional
generalization of unidirectional invisibility \cite{prsa-2016}. In
this letter, we use this formulation to solve the scattering problem
for the potentials of the form
    \be
    v(x,y)=\left[
    v_0(x)+ v_1(x)e^{i\alpha y}\right]\chi_a(x),
    \label{2d-potential}
    \ee
where $v_0(x)$ and $v_1(x)$ are piecewise continuous real- or
complex-valued functions, $v_0(x)\chi_a(x)$ is a potential whose
scattering problem in one dimension is exactly solvable, and
$\alpha$ and $a$ are positive real parameters. In particular, we
assume that the transfer matrix $\bM^{\rm(1d)}_{v_0}$ of
$v_0(x)\chi_a(x)$ in one dimension is given \cite{prl-2009,sanchez}.
Ref.~\cite{berry-98} considers the diffraction by absorbing crystals
that are modelled by a particular example of the potentials
(\ref{2d-potential}) with $v_0(x)$ and $v_1(x)$ being imaginary
constants.

We begin our analysis by recalling the definition of the transfer
matrix  in two dimensions. Let $v(x,y)$ be a potential such that for
$x\to\pm\infty$ the solutions of the Schr\"odinger equation,
    \be
    -\nabla^2\psi(x,y)+v(x,y)\psi(x,y)=k^2\psi(x,y),
    \label{sch-eq}
    \ee
tend to plane waves:
    \[\frac{1}{2\pi}\int_{-k}^k dp\, e^{ipy}\left[A_\pm(p)e^{i\omega(p)x}+
    B_\pm(p) e^{-i\omega(p)x}\right],\]
where $k$ is a positive real wavenumber, $A_\pm(p)$ and $B_\pm(p)$
are  coefficient functions vanishing for $|p|>k$, and
$\omega(p):=\sqrt{k^2-p^2}$. For a left-incident scattering solution
of (\ref{sch-eq}), $A_-(p)=2\pi\delta(p)$, $B_+(p)=0$, and
    \be
    \psi(r,\theta)\to e^{ikx}+\sqrt{i/kr}\,e^{ikr} f(\theta)~~{\rm as}~~ r\to\infty,
    \label{psi}
    \ee
where $(r,\theta)$ are polar coordinates, and $f(\theta)$ is the
scattering  amplitude of $v(x,y)$. The contribution of the
scattering potential to the reflected and transmitted waves are
respectively determined by $T_-(p):=B_-(p)$ and
$T_+(p):=A_+(p)-A_-(p)$. These by construction fulfil
    \be
    T_\pm(p)=0~~~~{\rm for}~~~~|p|>k.
    \label{condi}
    \ee

We identify the transfer matrix of the potential $v(x,y)$ with the
$2\times 2$ matrix operator $\bM(p)$ satisfying \cite{pra-2016}:
    \be
    \bM(p)\left[\begin{array}{c} A_-(p)\\ B_-(p)\end{array}\right]=
    \left[\begin{array}{c} A_+(p)\\ B_+(p)\end{array}\right].
    \label{M-def}
    \ee
Its entries, $M_{ij}(p)$, are linear operators that act in the space
$\sF$ of functions $\phi(p)$ vanishing for $|p|>k$.  $\bM(p)$ has
two basic properties: (i) It shares the composition property of its
one-dimensional analog \cite{sanchez}; (ii) It has complete
information about the scattering features of $v(x,y)$.

Property (i) follows from the fact that
    \be
    \bM(p)=\sT\exp\int_{-\infty}^\infty\!\!\!\!dx[-i\bH(x,p)],
    \label{M=}
    \ee
where $\sT$ stands for the time-ordering operation with $x$ playing
the  role of time, $\bH(x,p)$ is an effective non-Hermitian operator
given by
    \be
    \bH(x,p):=\frac{1}{2\omega(p)}\: e^{-i\omega(p)x\boldsymbol{\sigma}_3}
    v(x,i\partial_p)\,\boldsymbol{\cK}\,e^{i\omega(p)x\boldsymbol{\sigma}_3},
        \label{H=}
        \ee
$\boldsymbol{\sigma}_i$ are the Pauli matrices,
$\boldsymbol{\cK}:=\boldsymbol{\sigma}_3+i\boldsymbol{\sigma}_2$,
$v(x,i\partial_p)$ is the linear  operator acting in $\sF$ according
to
    \be
    v(x,i\partial_p)\phi(p):=\frac{1}{2\pi}\int_{-k}^k dq\,
    \tilde v(x,p-q) \phi(q),
    \label{v-dp}
    \ee
and $\tilde v(x,\fK_y)$ denotes the Fourier transform of $v(x,y)$
with respect to $y$, i.e., $\tilde v(x,\fK_y):=\int_{-\infty}^\infty
dy\,e^{-i\fK_y y}v(x,y)$.

Property (ii) is a result of the following relations \cite{pra-2016}.
    \begin{align}
    &T_-(p)=-2\pi M_{22}(p)^{-1}M_{21}(p)\delta(p),
        \label{Tm-L}\\
    &T_+(p)=M_{12}(p)T_-(p)+2\pi[M_{11}(p)-1]\delta(p),
    \label{Tp-L}\\
    &f(\theta)=-\frac{ik|\cos\theta|}{\sqrt{2\pi}}\times \left\{
    \begin{array}{cc}
    T_-(k\sin\theta) &{\rm for}~ \cos\theta<0\\
    T_+(k\sin\theta) &{\rm for}~  \cos\theta\geq 0
    \end{array}\right..
    \label{f=}
    \end{align}

The calculation of the effective Hamiltonian (\ref{H=}) for the
potentials  of the form (\ref{2d-potential}) is straightforward.
First, we substitute (\ref{2d-potential}) in (\ref{v-dp}) to show
that
    \be
    v(x,i\partial_p)=\chi_a(x)\left[v_0(x)+v_1(x)S_\alpha\right],
    \label{e1}
    \ee
where $S_\alpha$ is the shift operator defined by
    \be
    S_\alpha\phi(p):=\phi(p-\alpha).
    \label{shift}
    \ee
In view of (\ref{H=}) and (\ref{e1}),
    \begin{align}
    &\bH(x,p)=\bH_0(x,p)+\bH_1(x,p)S_\alpha,
    \label{H=HH}\\
    &\bH_j(x,p):=\frac{v_j(x)\chi_a(x)}{2\omega(p)}e^{-i\omega(p)x\bsigma_3}\bcK\,
    e^{i\omega(p-j\alpha)x\bsigma_3}.
    \label{Hj=}
    \end{align}%
where $j=0,1$. Because $\bH(x,p)$ vanishes for $x\notin[0,a]$, we
can replace  $-\infty$ and $\infty$ in (\ref{M=}) by $0$ and $a$,
respectively.

Next, we separate the contribution of $v_0(x)$ to the transfer
matrix $\bM(p)$ of  $v(x)$. To this end we introduce
    \begin{align}
    &\bM_0(x,p):=\sT\exp\int_{0}^x\!\!\!\!dx'[-i\bH_0(x',p)],
    \label{M-zero}\\
    &\widehat{\bH}_1(x,p):=\bM_0(x,p)^{-1}\bH_1(x,p)\bM_0(x,p-\alpha),
    \label{hat-H}\\
    &\widehat{\bM}(p):=\sT\exp\int_{0}^a\!\!\!\!dx[-i\widehat\bH_1(x,p)S_\alpha],
    \label{hat-M}
    \end{align}
so that $\bM_0(p):=\bM_0(\infty,p)=\bM_0(a,p)$ is the transfer
matrix of the  potential $v_0(x)\chi_a(x)$ in two dimensions, and
    \be
    \bM(p)=\bM_0(p)\widehat{\bM}(p).
    \label{M=MM}
    \ee
This reduces the calculation of $\bM(p)$ to that of $\bM_0(p)$ and $\widehat{\bM}(p)$.

The determination of $\bM_0(p)$ is equivalent to finding the
transfer matrix of  $v_0(x)\chi_a(x)$ in one dimension
\cite{pra-2014a} and substituting $\omega(p)$ for the wavenumber $k$
in the result;
    \be
    \bM_0(p)=\bM^{\rm(1d)}_{v_0}\Big|_{k\to\omega(p)}.
    \label{M-zero=2}
    \ee
In particular $\bM_0(p)$ is a known $2\times 2$ matrix with  unit
determinant. Notice also that
    \be
    \bM_0(x,p)=\bM_0(p)\Big|_{a\to x}=\bM^{\rm(1d)}_{v_0}\Big|_{k\to\omega(p),a\to x}.
    \label{M-x-p}
    \ee


To calculate  $\widehat{\bM}(p)$ we use (\ref{shift}) and the
definition of the  time-ordered exponential to  write (\ref{hat-M})
in the form
    \be
    \widehat{\bM}(p)=\bI+\sum_{n=1}^\infty \widehat{\bM}_n(p)S_{n \alpha}.,
    \label{series}
    \ee
Here $\bI$ is the $2\times 2$ identity matrix, and
    \begin{align}
    &\widehat{\bM}_n(p):=(-i)^n\!\!\!\int_{0}^a\!\!\!\!\!dx_n\!
    \int_{0}^{x_n}\!\!\!\!\!dx_{n-1}\! \cdots
    \int_{0}^{x_2}\!\!\!\!\!dx_1 \boldsymbol{\cH}_n ,
    \label{hat-Mn}\\
    &\begin{aligned}
    \boldsymbol{\cH}_n:=&
    \widehat\bH_1(x_n,p)\widehat\bH_1(x_{n-1},p-\alpha)\times\\
    &\widehat\bH_1(x_{n-2},p-2\alpha)\cdots\widehat\bH_1(x_1,p-(n-1)\alpha).
    \end{aligned}
    \label{Hn=}
    \end{align}

Now, consider an incident wave with wavenumber $k$ and
denote the integer part of $k/\alpha$ by $N(k)$. Then, according to
(\ref{shift}),  for any test function $\phi(p)$ that vanishes for
$p<-k$, $S_{[N(2k)+1]\alpha}\phi(p)=0$ for all $p$. Combining this
relation with the fact that $A_-(p)=B_-(p)=0$ for $p<-k$ and using
(\ref{M-def}) and  (\ref{M=MM}), we see that the terms on the
right-hand side of (\ref{series}) vanish for $n>N(2k)$, i.e.,
$\widehat{\bM}(p)=\bI+\sum_{n=1}^{N(2k)}
\widehat{\bM}_n(p)S_{n\alpha}$. As a result, (\ref{M=MM}) gives
    \bea
    {\bM}(p)&=&\sum_{n=0}^{N(2k)}{\bM}_n(p)S_{n \alpha},
    \label{M=Sum}
    \eea
where
    \be
    {\bM}_n(p):=\bM_0(p)\widehat{\bM}_n(p)~~~{\rm for}~~~n\geq 1.
    \label{Mn=def}
    \ee
According to (\ref{M=Sum}) the entries of ${\bM}(p)$ have the form:
    \be
    M_{ij}(p)={M}_{ij}^{(0)}(p)+
    \sum_{n=1}^{N(2k)}{M}_{ij}^{(n)}(p)S_{n\alpha},
    \label{Mij-=}
    \ee
where ${M}_{ij}^{(n)}(p)$ are the entries of  $\bM_n(p)$. It is
important to note that these are functions of $p$ that act on other
functions of $p$ by multiplication.

In order to determine the scattering amplitude $f(\theta)$ for the
potentials  (\ref{2d-potential}) we compute the reflection and
transmission functions $T_\pm(p)$. To do this we first write
(\ref{Tm-L}) in the form
    \be
    M_{22}(p)T_-(p)=-2\pi M_{21}(p)\delta(p),
    \label{Tm-L-n}
    \ee
and propose the following ansatz for its solution:
    \bea
    T_-(p)=2\pi\sum_{n=0}^{N(k)} t_{n}^{-}\,\delta(p-n\alpha),
    \label{anstaz}
    \eea
where
$t_{n}^{-}$ are coefficients that we determine by inserting
(\ref{anstaz}) in  (\ref{Tm-L-n}) and making use of (\ref{Mij-=})
and (\ref{condi}). The result is the following iterative
construction of $t_{n}^{-}$.
    \begin{align}
    t_{0}^{-}=&-\frac{M_{21}^{(0)}(0)}{M_{22}^{(0)}(0)},
    \label{t-zero}\\
    t_{\ell}^{-}=&-\frac{M_{21}^{(\ell)}(\ell\alpha)+\sum_{m=1}^\ell
    M_{22}^{(m)}(\ell\alpha)\,t_{\ell-m}^-}{M_{22}^{(0)}(\ell\alpha)},
    \label{t-ell}
    \end{align}
where $\ell=1,2,\cdots,N(k)$. Remarkably, $t_{0}^{-}$ coincides with
the left reflection  amplitude $R^{\rm l}$ for the potential
$v_0(x)\chi_a(x)$ in one dimension \cite{prl-2009}.

We can view (\ref{t-zero}) and (\ref{t-ell}) as a linear system of
$N':=N(k)+1$ equations  for $t_n^{-}$. Identifying the latter with
the entries of a column vector $\bt^-$, we can write the solution of
these equations as $\bt^-=\sum_{n=0}^{N(k)} \bA^{\!n}\,\bb$, where
$\bA$ and $\bb$ are respectively the $N'\times N'$ and $N'\times 1$
matrices with entries:
    \bea
    A_{nn'}&:=&\left\{\begin{array}{cc}
    -\frac{M_{22}^{(n-n')}(n\alpha)}{M_{22}^{(0)}(n\alpha)}
    &{\rm for}~0\leq n'\leq n-1,\\
    0&{\rm otherwise},\end{array}\right.\nn\\
    b_n&:=&-\frac{M_{21}^{(n)}(n\alpha)}{M_{22}^{(0)}(n\alpha)},~~~~
    n,n'=0,1,2,\cdots N(k).\nn
    \eea

Next, we use (\ref{anstaz}) to write (\ref{Tp-L}) in the form
    \be
    T_+(p)=2\pi\sum_{n=0}^{N(k)} t^+_n\,\delta(p-n\alpha),
    \label{T-plus=}
    \ee
where
    \bea
    t^+_0&:=&M_{12}^{(0)}(0)\,t^-_0+M_{11}^{(0)}(0)-1=
    \frac{1}{M_{22}^{(0)}(0)}-1,~~~~
    \label{t-zero-p}\\
    t^+_\ell&:=&M_{11}^{(\ell)}(\ell\alpha)+
    \sum_{m=0}^\ell M^{(\ell-m)}_{12}(\ell\alpha)\,t^-_m.
    \label{t-ell-p}
    \eea
The second equality in (\ref{t-zero-p}) follows from the fact that
$\det\bM_0(0)=1$. Note also  that $t^+_0$ coincides with $T-1$ where
$T$ is the transmission amplitude for the potential
$v_0(x)\chi_a(x)$ in one dimension \cite{prl-2009}.

Having computed $T_\pm(p)$, we use (\ref{f=}) to obtain the
following expression for the  scattering amplitude:
    \begin{align}
    &f(\theta)=-i\sqrt{2\pi}\sum_{n=0}^{N(k)}\left[
    t^+_n\,\delta(\theta-\theta_{n}^+)+t^-_n\,
    \delta(\theta-\theta_{n}^-)\right],
    \label{f=exp}
    \end{align}
where $\theta\in[-\pi/2,3\pi/2)$ and
    \be
    \theta_n^+:=\arcsin\mbox{$\left(\frac{n\alpha}{k}\right)$}
    \in [0,\mbox{$\frac{\pi}{2}$}],~~~
    \theta_n^-:=\pi-\theta_n^+\in [\mbox{$\frac{\pi}{2}$},\pi].
    \label{theta=}
    \ee
This completes the solution of the scattering problem for the
potentials (\ref{2d-potential}).  To summarize, we list its basic
steps:
    \begin{enumerate}
    \item For a given $k$, determine $N(k)$
    and do the following for all $n=0,1,2,\cdots,N(k)$;
    \item Calculate $\bH_1$, $\widehat{\bH}_1$, $\boldsymbol{\cH}_n$,
    and $\widehat{\bM}_n(p)$
    using (\ref{Hj=}), (\ref{hat-H}), (\ref{Hn=}), and (\ref{hat-Mn});
    \item Find $\bM_n(p)$ and its entries $M_{ij}^{(n)}(p)$ using (\ref{Mn=def});
    \item Determine $t^\pm_n$ and $f(\theta)$ using (\ref{t-zero}) -- (\ref{f=exp}).
    \end{enumerate}

The $n=0$ term on the right-hand side of (\ref{f=exp}) gives the
contribution of  $v_0(x)$ to the scattering amplitude. For
$k<\alpha$, this is the only term contributing to $f(\theta)$.
Therefore, the presence of $v_1(x)$ does not affect the scattering
process. In particular, a potential of the form $v_1(x)e^{i\alpha
y}\chi_a(x)$ does not scatter any incident plane wave with
wavevector pointing along positive $x$-axis and $k<\alpha$, i.e., it
is invisible for these waves.



For $k\geq\alpha$, the scattered wave consists of $N(k)$ pairs of
plane waves $\psi_n^\pm$ propagating along the direction given by
$\theta_n^\pm$; their wavevector has the form:
$(\pm\sqrt{k^2-(n\alpha)^2},n\alpha)$, \cite{footnote-01}. In
particular, they have the same $y$-component which is quantized in
units of $\alpha$, while their $x$-component are equal in magnitude
and opposite in sign. This observation, which might find
applications in quantum metrology, also applies to wave packet
scattering.

Consider an incident wave packet whose momentum is directed along
the $x$-axis and has a spread  about its average value that is much
smaller than $\alpha$. The scattering of such a wave packet by a
potential of the form (\ref{2d-potential}) leads to scattered
wave-packet pairs that are sharply peaked in the momentum space. The
$y$-component of the momentum of the members of each pair are
identical and quantized in units of $\alpha$, while their
$x$-component add up to zero. These correspond to a quantum state
with entangled momentum. Potentials (\ref{2d-potential}) may,
therefore, serve as sources for entangled states.

The scattering features of the potentials (\ref{2d-potential}) is
reminiscent of the photoelectric effect. The incident wave and the
resulting pairs of scattered waves respectively play the role of an
incident photon on a metallic plate and the emitted electrons by the
plate. Similarly to the photoelectric effect, a scattered wave
appears only if the energy of the incident wave exceeds a critical
value.



Next we examine a simple optical realization of the potentials of
the  form (\ref{2d-potential}). Consider an infinite optical slab
placed between the planes $x=0$ and $x=a$ in three-dimensions, and
suppose that its relative permittivity $\hat\varepsilon$ does not
depend on $z$. Let $\bbe_j$ be the unit vector pointing along the
$j$-axis for $j=x,y,z$. The scattering of a normally incident
$z$-polarized TE wave, $E_0 e^{ik(x-c t)}\bbe_z$, by this slab is
described by the Schr\"odingier equation (\ref{sch-eq}) for the
optical potential: $v(x,y)=k^2[1-\hat\varepsilon(x,y)]$,
\cite{prsa-2016}. If this potential has the form
(\ref{2d-potential}), the scattering problem for these TE waves is
equivalent to the one we have treated above. In particular, we can
use (\ref{psi}) and (\ref{f=exp}) to derive the following asymptotic
expression for the time-averaged Poynting vector for the scattered
TE waves \cite{prsa-2016}:
     {\small \be
     \br\bS\kt=\frac{|E_0|^2}{2\mu_0 {\rm c}}\left\{
     \begin{aligned}
     &(1+|t^-_0|^2){{\bbe}}^-_0\!
     +\!\sum_{n=1}^N\left|t^-_n\right|^2{{\bbe}}_n^-
     ~~{\rm for}~x\to-\infty,\\[-3pt]
     &\left|1+t_0^+\right|^2{{\bbe}}^+_0\!+\!\sum_{n=1}^N\left|t^+_n\right|^2{{\bbe}}^+_n
     ~~{\rm for}~x\to\infty,
     \end{aligned}
     \right.
     \nn
     \ee}%
where $\mu_0$ and ${\rm c}$ are respectively the permeability and
speed of light in vacuum, and
$\bbe^\pm_n:=\cos\theta_n^\pm\bbe_x+\sin\theta_n^\pm\bbe_y$.

The slab we have introduced involves a particular periodic
modulation  along the $y$-direction, but there is no restriction on
the choice of $v_0(x)$ and $v_1(x)$, except that $v_0(x)\chi_a(x)$
must be exactly solvable in one dimension. This is a great
advantage. In contrast to the unidirectionally invisible systems
\cite{lin,invisible2} modeled by the potentials of the form
(\ref{1d-potential}), which require balanced regions of gain and
loss, we can consider a lossy slab by taking $v_0(x)$ to have a
positive imaginary part larger than $|v_1(x)|$. For example, let
$v_j(x)=-k^2\fz_j$, where $\fz_j$ are complex numbers such that
$0<|\fz_1|<\IM(\fz_0)$. Then
    \be
    \hat\varepsilon(x,y)=1+(\fz_0+\fz_1\,e^{i\alpha y})\chi_a(x),
    \label{epsilon-slab}
    \ee
and the slab does not involve any gain regions. This is reminiscent
of the concept of passive $\cP\cT$-symmetry \cite{gao}.

For an incident wave with $k\in(\alpha,2\alpha)$, $N=1$ and the
scattered wave is determined in terms of $t_0^\pm$ and $t_1^\pm$.
For a slab described by the permittivity
profile~(\ref{epsilon-slab}) we compute these in the appendix. The result is
    \be
    t_1^\pm=\frac{\fz_1 k^3 A^\pm}{\fn^2\alpha^2\omega_1 B},
    \ee
where $\fn:=\sqrt{1+\fz_0}$ is the refractive index of the
unmodulated slab,
$\omega_n:=\sqrt{k^2-(n\alpha)^2}=k\cos\theta^+_n$,
    \bea
    A^-&:=& -k^{-1}\big\{k_-[1-\cos(ak_+\fn)]+\nn\\
    &&i[\fn_-k_+\sin(ak_-\fn)+\fn_+k_-\sin(ak_+\fn)]\big\}\nn\\
    A^+&:=&k^{-1}e^{-ia\omega_1}\big\{k_+[\cos(ak\fn)-\cos(a\omega_1\fn)]+\nn\\
    &&i[(\fn^{-1}k+\fn\omega_1)\sin(a\omega_1\fn)-
    (\fn k+\fn^{-1}\omega_1)\sin(ak\fn)]\big\}\nn\\
    B&:=&\fn_-^2\!\cos(ak_-\fn)-(\fn_+^2\!+1)\!\cos(ak_+\fn)+
    2i\fn_+\!\sin(ak_+\fn),\nn
    \eea
$k_\pm:=k\pm\omega_1$, and $\fn_\pm=(\fn\pm\fn^{-1})/2$.

\begin{figure}
    \begin{center}
    \includegraphics[width=.2\textwidth]{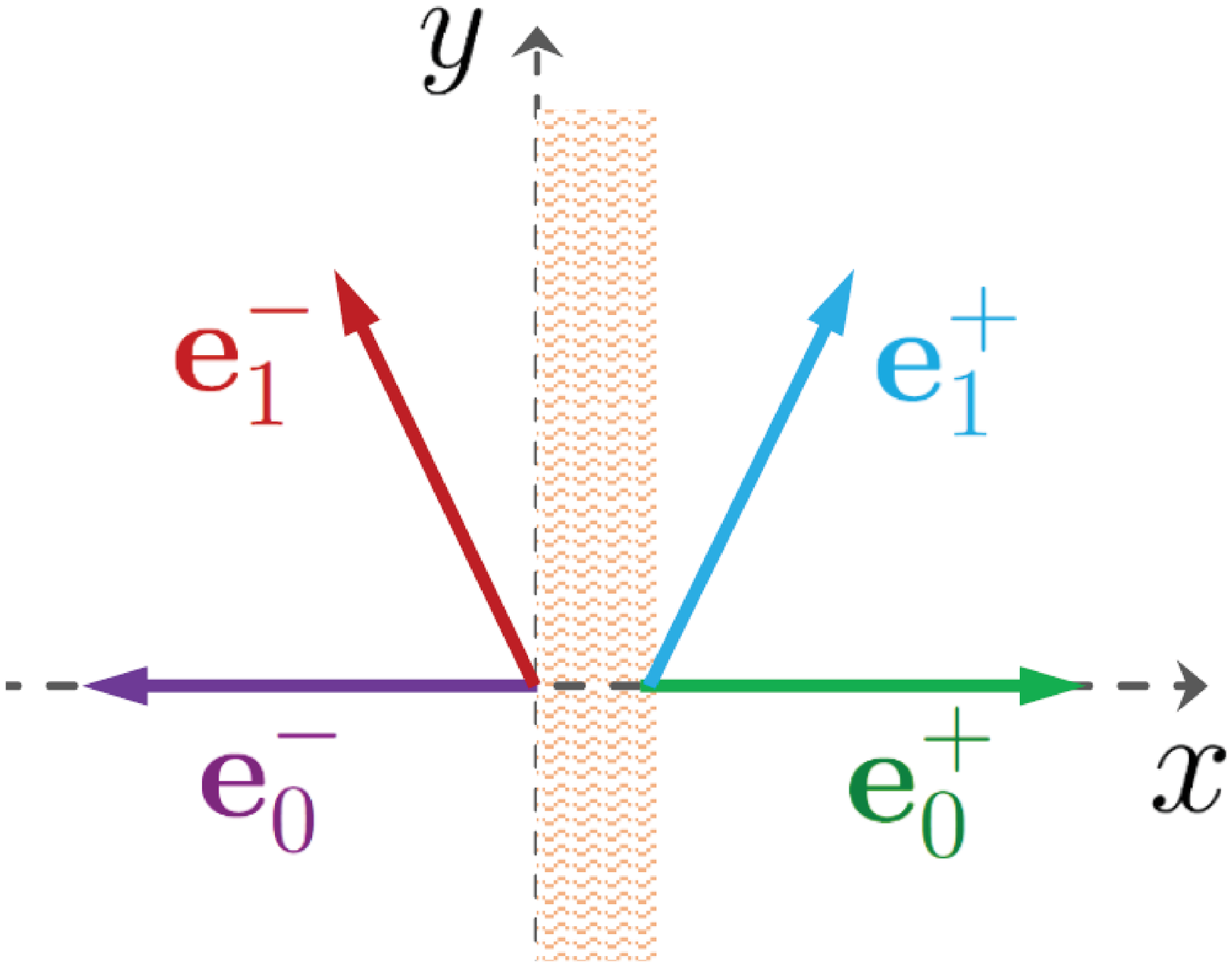}~~~
    \includegraphics[width=.23\textwidth]{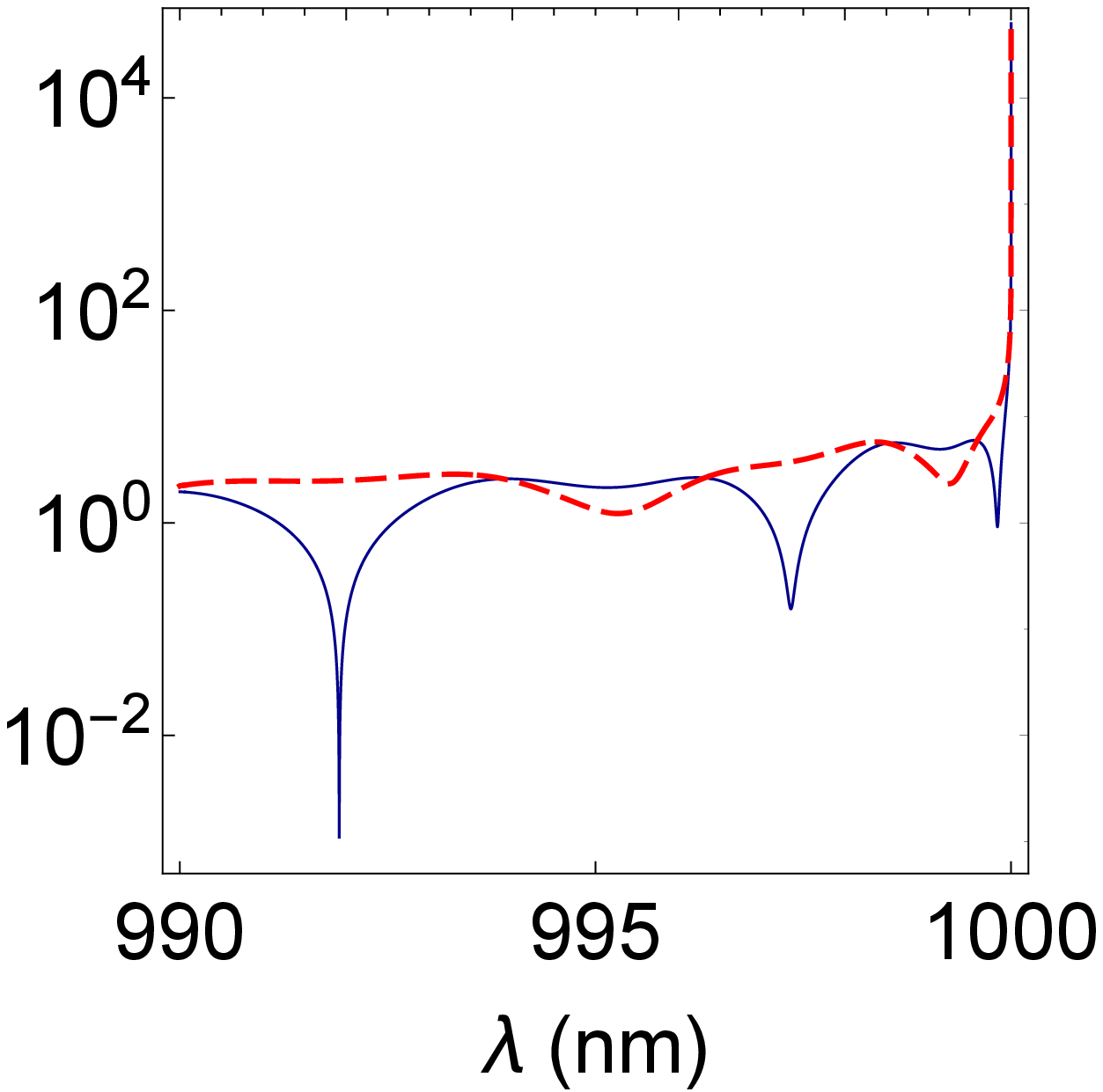}\vspace{-6pt}
    \caption{Left-hand figure shows a schematic view of the directions $\bbe^\pm_n$ along which
    a left-incident wave with $k\in(\alpha,2\alpha)$ scatters by a slab
    described by the permittivity profile~(\ref{epsilon-slab}). Right-hand figure
    shows the graphs of $t^-_1/\fz_1$ (solid navy curve) and $t^+_1/\fz_1$ (red dashed curve) as
    functions of the wavelength $\lambda$ for a lossy
    Titanium-Sapphire slab with $a=10~\mu{\rm m}$, $\fn=2.892+i(4.5\times 10^{-5})$,  and
    $\lambda_\alpha=2\pi/\alpha=1000~{\rm nm}$. For the selected range of values of
    $\lambda$, $\theta_1^+\in[82^\circ,90^\circ]$.}
    \label{fig1}
    \end{center}
    \end{figure}

Figure~\ref{fig1} shows the plots of $\fz_1^{-1}|t^\pm_1|$ as
functions of the wavelength $\lambda:=2\pi/k$ for a
Titanium-Sapphire slab with thickness $a=10\,\mu{\rm m}$ and
modulation period $\lambda_\alpha:=2\pi/\alpha=1~\mu{\rm m}$.
As seen from this figure, the critical wavelength
$\lambda=\lambda_\alpha$ is a singularity of $|t^\pm_1|$. It is
indeed easy to show that for $k=\alpha+\delta k$ with $0<\delta k\ll
\alpha$, $\theta_1^\pm\approx 90^\circ\mp\delta\theta$ where
$\delta\theta:=\sqrt{2\delta k/\alpha}\ll 1$, and $t^\pm_1\approx
     -\fz_1\{\fn[\fn+i\cot(a\alpha\fn/2)]\delta\theta\}^{-1}$.
This shows that the oblique scattered waves tend to have a much
larger amplitude when the wavelength of the incident wave is close
to the period of the modulation $\lambda_\alpha$. The same holds for
the scattered waves $\psi_N^\pm$ when $\lambda$ approaches
$\lambda_\alpha/N$ for some positive integer $N$. This is a novel
resonance effect that should be experimentally observable.
Remarkably it is present also for a purely lossy slab. It is easy to
show that this effect has  its root in the fact that in general
$t_n^\pm=\fz_1^n\alpha\hat t_n^\pm/\omega_n$ where $\hat t_n^\pm$
take a nonzero value for $k\approx n\alpha$. For values of $k$ close
to $N\alpha$,  $\omega_N\ll \alpha$ which implies that $|t_N^\pm|\gg
|\fz_1|^{n-N}|t_n^\pm|$ for $n<N$.

Next, suppose that the slab contains gain regions. We can arrange
its  parameters so that for some $k>n\alpha$ and $n\geq 1$,
$M_{22}^{(0)}(n\alpha)$ vanishes and $|t^\pm_n|$ blow up. This
corresponds to the emergence of a spectral singularity
\cite{prl-2009}. It signals the onset of lasing in the TE mode of
the slab with incidence (emission) angle $\theta_n^\pm$. For the
permittivity profile~(\ref{epsilon-slab}), we can easily compute
$M_{22}^{(0)}(n\alpha)$ and impose $M_{22}^{(0)}(n\alpha)=0$ to
obtain the threshold gain $g_\star$ and the lasing modes $k_\star$,
\cite{pra-2015a}. For typical high gain material, where
$|\IM(\fn)|\ll\RE(\fn)-1\ll a k_\star$, this gives
    \bea
    g_\star&=&\frac{4\sqrt{\eta^2-\sin^2\theta_n^+}}{\eta a}
    \ln\!\!\left[\frac{\sqrt{\eta^2-\sin^2\theta_n^+}+|\cos\theta_n^+|}{
    \sqrt{\eta^2-1}}\right]\!\!,~~~~
    \label{g=}\\
    k_\star&=&\sqrt{(\pi m/\eta a)^2+(\alpha n)^2}=
     n\alpha/\sin\theta_n^+,
    \label{k=}\\
    \theta_n^+&=&{\rm arcsin}\left[1/\sqrt{1+ ( \pi q/\eta \alpha a)^2}\right]\!\!,
    \label{theta-plus=}
    \eea
where $\eta:=\RE(\fn)$, $m=1,2,3,\cdots$ is a mode number, $q:=m/n$,
and we have used (\ref{theta=}).

According to (\ref{theta=}) -- (\ref{theta-plus=}), $\theta_n^\pm$
and  consequently $g_\star$ depend on $q$, while $k_\star$ is a
function of both $n$ and $r$. To make this more transparent we use
$\theta^\pm(q)$, $g_\star(q)$, and $k_n(q)$ for $\theta_n^\pm$,
$g_\star$, and $k_\star$, respectively. If we attain the threshold
gain $g_\star(q)$ for a particular value of $q$, the slab will begin
emitting laser light in the directions given by $\theta^\pm(q)$ and
wavenumbers $k_n(q)=n k_1(q)$ for different values of $n$, i.e., it
functions as a directional laser with a degenerate set of lasing
modes. Clearly, this argument ignores the effects of dispersion and
does not ensure the presence of the nonlinear effects that are
necessary for having stable laser oscillators. Yet it reveals the
practical importance of the longitudinal mode number $n$ that stems
from the periodic modulation along the $y$-direction. This is an
active analog of the grating component of the distributed feedback
lasers. But unlike the latter its role is to produce an additional
mode number that might find interesting applications in designing
realistic multimode lasers \cite{ge-2017}.

The discovery of a new exactly solvable class of potentials is a
very  rare event in the history of quantum physics. Here we report
such a discovery and discuss some of its potential applications. In
particular, we draw attention to the relevance of our findings for
the production of entangled states with a quantized $y$-component of
momentum. We also outline a simple optical realization of our
potentials and elaborate on their potential application in
generating degenerate lasing modes. We also reveal a resonance
effect arising for wavenumbers that are integer multiples of
$\alpha$. This leads to scattering waves of large amplitude
traveling at grazing angle. It turns out that the singular behavior
of these waves is regularized when one considers a slab with a
finite extension along the $y$-axis. A detailed study of this effect
is a subject of future investigation.

Our results can be easily generalized to the treatment of scattering
of oblique incident waves. As explained in \cite{p133}, this amounts
to replacing $\delta(p)$ in (\ref{Tm-L}) and (\ref{Tp-L}) by
$\delta(p-p_0)$, where $p_0$ is the $y$-component of the wavevector
for the incidence wave. Another possible extension is the solution
of the scattering problem for the potentials of the form
$v(x)=[v_0(x)+\sum_{j}v_j(x)e^{i\alpha_jy}]\chi_a(x)$ with
$\alpha_j>0$.\vspace{6pt}

\noindent {\em Acknowledgments.}--- We are indebted to Li Ge, Vahid
Karimipour, Hamid-Reza Ramazani, Alphan Sennaro\u{g}lu, and Ali
Serpeng\"{u}zel for fruitful discussions. This project was supported
by Turkish Academy of Sciences (T\"UBA).

\np

\section*{Appendix}

In the following we give details of the calculation of $t_0^\pm$.

According to Eqs.~(31), (32), (34), and
(35),
    \begin{align}
    &t_0^-=-\frac{M_{21}^{(0)}(0)}{M_{22}^{(0)}(0)},
    &&t_0^+=\frac{1}{M_{22}^{(0)}(0)}-1,
    \label{sm-t0}
    \end{align}
    \be
    \begin{aligned}
    &t_1^-=-\frac{M_{21}^{(1)}(\alpha)+M_{22}^{(1)}(\alpha)t_0^-
    }{M_{22}^{(0)}(\alpha)},\\
    &t_1^+=M_{11}^{(1)}(\alpha)+M_{12}^{(1)}(\alpha)t_0^-+
    M_{12}^{(0)}(\alpha)t_1^-,
    \end{aligned}
    \label{sm-t1}
    \ee
where $M_{ij}^{(n)}(p)$ are the entries of $\bM_n(p)$.

For the permittivity profile (38),
    \bea
    v_j(x)&=&-k^2\fz_j\chi_a(x),~~~~j=0,1.
    \label{vj=}
    \eea
In particular the expression for $\bM_0(p)$ follows from the formula
for the transfer matrix of a homogenous slab. As shown in [F.~Loran
and A.~Mostafazadeh, Phys.\ Rev.~A \textbf{93}, 042707 (2016)], it
has the form:
    \be
    \bM_0(p)=\left[\begin{array}{cc}
    F_0(a,\omega) & G_0(a,\omega)\\
    G_0(a,-\omega) & F_0(a,-\omega)\end{array}\right],
    \label{M-zero=gh}
    \ee
where
    \bea
    &&F_0(a,\omega):=\left[\cos(a\,\omega\fn)+i\fn_+\sin(a\,\omega\fn)\right]
e^{-ia\omega},\nn\\
    &&G_0(a,\omega):=i\fn_-\sin(a\,\omega\fn)e^{-ia\omega},\nn\\
    &&\omega:=\omega(p):=\sqrt{k^2-p^2},\nn\\
    &&\fn:=\sqrt{1+\fz_0},~~~~\fn_\pm:=\frac{\fn^2\pm 1}{2\fn}.\nn
    \eea
In view of (\ref{M-zero=gh}),
	\be
    \begin{aligned}
    &M_{11}^{(0)}(\alpha)=F_0(a,\omega_1),
    &&M_{12}^{(0)}(\alpha)=G_0(a,\omega_1),\\
    &M_{21}^{(0)}(\alpha)=G_0(a,-\omega_1),
    &&M_{22}^{(0)}(\alpha)=F_0(a,-\omega_1),
    \end{aligned}
    \label{MM}
    	\ee
where
	\[\omega_1:=\omega(\alpha)=\sqrt{k^2-\alpha^2}=k\cos\theta^+_1.\] 
Eqs.~(\ref{MM}) together with (\ref{sm-t0}) imply
    \begin{align}
    &t^-_0=-\frac{G_0(a,-k)}{F_0(a,-k)},
    &&t^+_0=\frac{1}{F_0(a,-k)}-1.
    \label{t-minuses}
    \end{align}

In order to compute $t_1^\pm$ we need to determine $\bM_1(\alpha)$. First, we recall
that according to Eqs.~(27), (24), (25),
(18), (22), (\ref{M-zero=gh}), and (16),
    \begin{align}
    \bM_1(\alpha)&=\bM_0(\alpha)\widehat\bM_1(\alpha),
    \label{sm-e2}\\
    \widehat\bM_1(\alpha)&=-i\int_0^a dx\: \widehat{\bH}_1(x,\alpha),
    \label{sm-e3}\\
    \widehat{\bH}_1(x,\alpha)&=\bM_0(x,\alpha)^{-1}\bH_1(x,\alpha)\bM_0(x,0),
    \label{sm-e4}\\
    \bM_0(x,\alpha)&=\left[\begin{array}{cc}
    F_0(x,\omega_1) & G_0(x,\omega_1)\\
    G_0(x,-\omega_1) & F_0(x,-\omega_1)\end{array}\right],
    \label{sm-e5}
    \end{align}
    \vspace{-15pt}
    \bea
    \bH_1(x,\alpha)&=&\frac{v_1(x)\chi_a(x)}{2\omega_1}e^{-i\omega_1 x\bsigma_3}
	\bcK\, e^{ikx\bsigma_3}\nn\\
	&=&\frac{v_1(x)\chi_a(x)}{2\omega_1}\left[\begin{array}{cc}
    	e^{i k_-x} & e^{-ik_+ x}\\
    	-e^{ik_+x}& -e^{-ik_-x}\end{array}\right],
	\label{sm-e6}
    \eea
where
    \[k_\pm:=k\pm\omega_1.\]
Next, we substitute (\ref{sm-e5}) and (\ref{sm-e6}) in (\ref{sm-e4}) and
simplify the resulting expression to obtain
    \be
    \widehat{\bH}_1(x,\alpha)=\frac{v_1(x)\chi_a(x)}{2\omega_1\fn}
    \left[\begin{array}{cc}
    h(k_+,k_-) & h(-k_-,-k_+)\\
    -h(k_-,k_+) & -h(-k_+,-k_-)\end{array}\right],
    \label{sm-e8}
    \ee
where
    \bea
    &&h(\mu,\nu):=\fn_+\cos(\nu\fn x)+\fn_-\cos(\mu\fn x)
    +i\sin(\nu\fn x).\nn
    \eea

In view of (\ref{vj=}) and (\ref{sm-e8}) we can evaluate the
integral in (\ref{sm-e3}). Substituting the result in
(\ref{sm-e2}) and using (\ref{M-zero=gh}), we obtain
    \be
    \bM_1(\alpha)=\left[\begin{array}{cc}
    F_1(k,\omega_1)&F_1(-k,\omega_1)\\
    F_1(k,-\omega_1)&F_1(-k,-\omega_1)\end{array}\right],
    \label{e1}
    \ee
where
    \begin{align}
    F_1(\mu,\nu):=&\frac{\fz_1  \mu^2 e^{-ia\nu}}{2\fn^3(\mu^2-\nu^2)\nu}
    \Big\{\fn(\mu+\nu)[\cos(a\mu\fn)-\cos(a\nu\fn)]+
    \nn\\
    &i\left[(\mu\fn^2+\nu)\sin(a\mu\fn)-
    (\mu+\nu\fn^2)\sin(a\nu\fn)\right]\Big\}.\nn
    \end{align}
According to (\ref{e1}),
	\be
    \begin{aligned}
    &M_{11}^{(1)}(\alpha)=F_1(k,\omega_1),
    &&M_{12}^{(1)}(\alpha)=F_1(-k,\omega_1),\\
    &M_{21}^{(1)}(\alpha)=F_1(k,-\omega_1),
    &&M_{22}^{(1)}(\alpha)=F_1(-k,-\omega_1).
    \end{aligned}
    	\label{last}
	\ee
If we insert (\ref{t-minuses}) and (\ref{last}) in (\ref{sm-t1}), we find that $t_1^\pm$ 
are given by (39).

\ed

\ed

The expression for $\bM_0(p)$ follows from the formula for the
transfer matrix of a homogenous slab \cite{pra-2016}. It has the
form
    \be
    \bM_0(p)=\left[\begin{array}{cc}
    F_0(\omega) & G_0(\omega)\\
    G_0(-\omega) & F_0(-\omega)\end{array}\right],
    \label{M-zero=gh}
    \ee
where
$F_0(\omega):=\left[\cos(a\omega\fn)+i\fn_+\sin(a\omega\fn)\right]
e^{-ia\omega}$, $G_0(\omega):=i\fn_-\sin(a\omega\fn)e^{-ia\omega}$,
and $\omega:=\sqrt{k^2-p^2}$. Substituting (\ref{M-zero=gh}) in
(\ref{t-zero}) and (\ref{t-zero-p}), we have
    \begin{align}
    &t^-_0=-\frac{G_0(-k)}{F_0(-k)},
    &&t^+_0=\frac{1}{F_0(-k)}-1.
    \label{t-minuses}
    \end{align}
The computation of $\bM_1(p)$ is more involved but the result is
surprisingly simple:
    \begin{align}
    &\bM_1(p)=\left[\begin{array}{cc}
    F_1(k,\omega)&F_1(-k,\omega)\\
    F_1(k,-\omega)&F_1(-k,-\omega)\end{array}\right],
    \label{M1=}\\
    &
    \begin{aligned}
    F_1(k,\omega):=&\frac{\fz_1 k^2 e^{-ia\omega}}{2\fn^3(k^2-\omega^2)\omega}
    \Big\{\fn(k+\omega)[\cos(ak\fn)-\cos(a\omega\fn)]
    \nn\\
    &\hspace{-12pt}+i\left[(k\fn^2+\omega)\sin(ak\fn)-
    (k+\omega\fn^2)\sin(a\omega\fn)\right]\Big\}.
    \end{aligned}\nn
    \end{align}


\begin{thebibliography}{99}
\bibitem{footnote-01} It is easy to see that $\theta_n^\pm$ satisfy $\phi_n<\theta_n^+\leq \pi/2\leq\theta_n^-<\pi-\phi_n$, where $\phi_n:=\arcsin(n/[N(k)+1])$. Therefore the scattered plane-wave pairs propagate along directions that lie in the wedge-shaped region $\phi_1<\theta<\pi-\phi_1$.
\end{thebibliography}
\end{document}